\documentclass[12pt]{article}

\usepackage{epsfig}

\begin{document}

\title{Teleportation: from probability distributions to quantum
  states} 

\author{ M. Koniorczyk $^{\dag,\ddag}$, T. Kiss $^{\dag,\ddag}$,
  J. Janszky$^\dag$\\
\small $^\dag$\ Department of Nonlinear and Quantum Optics, Research\\
\small  Institute for Solid State Physics and Optics, Hungarian Academy of\\
\small  Sciences, P.O. Box 49, H-1525 Budapest, Hungary\\
\small $\ddag$\ Institute of Physics, University of P\'ecs,\\
\small  Ifj\'us\'ag \'utja 6. H-7624 P\'ecs, Hungary}

\maketitle

\begin{abstract}
  
  The role of the off-diagonal density matrix elements of the entangled pair
  is investigated in quantum teleportation of a qubit. The dependence between
  them and the off-diagonal elements of the teleported density matrix is shown
  to be linear. In this way the ideal quantum teleportation is related to an
  entirely classical communication protocol: the one-time pad cypher. The
  latter can be regarded as the classical counterpart of Bennett's quantum
  teleportation scheme. The quantum-to-classical transition is demonstrated on
  the statistics of a gedankenexperiment.\\
{PACS Nubmers: 03.67.Hk, 03.65.Bz}
\end{abstract}

\section{Introduction}
The quantum teleportation phenomenon \cite{prl70_1895} has become one of the
central aspects in the investigations of quantum entanglement.  Its possible
applicability for quantum communication and its experimental realization in
different physical systems
\cite{nature390_575,prl80_1121,science282_706,nature396_52} gave rise to a
wide spread interest in several related problems, including possible
alternative schemes
\cite{pra60_2759,pra58_4373,pra62_013802,prl84_3486,pla272_32}, discussion and
resolution of experimental and theoretical limitations
\cite{pla273_158,pra59_3295,pra61_042304} and the connection to nonlocality and
inseparability \cite{prl72_797,pla210_157,pra60_1888}.

The relation of nonlocality and inseparability of mixed states has already
been a challenging problem before the advent of quantum teleportation. Werner
\cite{pra40_4277} introduced inseparable mixed states which do not violate
Bell's inequalities.  Several measures have been defined to quantify the
amount of entanglement contained in mixed states
\cite{pra57_1619,prl80_2245,pra54_3824,pra59_141}, partly inspired by
entanglement distillation protocols
\cite{pra53_2046,pla210_151,prl76_722,pra61_032305}.  Whether it is possible
to employ a bipartite mixed state for quantum teleportation has turned out to
be an additional nontrivial question in this field
\cite{prl72_797,quantph0008038,pra62_032305}.  Recently, Bose and Vedral
\cite{pra61_040101} have investigated the direct relation between the
mixedness of a state and its usefulness for teleportation.  They applied the
von Neumann entropy $ S(\varrho)=-\mbox{Tr} (\varrho \ln \varrho)$ as 
a measure for
mixedness, and find that there is a threshold for this quantity above which
the bipartite state $\varrho$ becomes useless for teleportation.

In order to separate the quantum and classical aspects of communication, N.
and B. Gisin \cite{pla260_323} and independently Steiner \cite{quantph9902014}
have presented a local hidden variable (LHV) model which, supplied by
classical communication, is capable of reproducing quantum correlations of two
spin 1/2 particles in singlet state.  Brassard et al.  \cite{prl83_1874}
quantify the amount of classical communication required to be supplemented to
simulate quantum correlations for given numbers of qubits.  More recently, Cerf
et al. \cite{prl84_2521} have presented a protocol in an LHV model for
teleporting a quantum state via classical communication. On the other hand, M.
\.  Zukowski \cite{pra62_032101} has investigated the connection of local
realism and the nonclassical part quantum teleportation process, and
concluded, that no local hidden variable model can reproduce the quantum
prediction.

The effect of mixedness of a bipartite state applied for teleportation is
obviously an important aspect in understanding the classical--quantum limit.
The classical limit of the original protocol of Bennett has not yet been
investigated. Quantum teleportation, the basic primitive of quantum
communication networks consists of the transmission of a quantum state via a
combination of classical channels and quantum correlations. In lack of either
the classical channel or the entangled resource, only noise is obtained. But
what is the classical analogue of Bennett's scheme? To answer this question,
we investigate a set of bipartite states interpolating between an ideal
entangled state and a possible classical limit.  This approach provides a
transition from the quantum teleportation to its classical counterpart.

In this paper we examine, what happens to a teleportation scheme, when the
density matrix elements of the entangled state used as a correlated pair,
which are off-diagonal on product basis, are reduced. This approach, turning
the entangled state into a classical correlation, obviously offers a possible
way of obtaining a classical limit of the teleportation process.

The paper is organized as follows: in section \ref{sec:telep} the class of
bipartite states in argument is described, and Bennett's protocol of quantum
teleportation of a qubit is summarized in a consistent density matrix
formalism.  The latter can be regarded as a special case of the treatment of
e. g. \cite{pra44_2547} or \cite{pra60_1888}. By replacing the ideal EPR pair
with the states investigated, we obtain our main result.  Starting from this,
two examples are studied in detail: in section \ref{sec:class} the purely
classical limit is introduced, and in section \ref{sec:stat} cases between the
ideal quantum teleportation and the classical limit are analyzed by examining
a gedankenexperiment. Section \ref{sec:concl} summarizes the results.

\section{Quantum teleportation revisited}
\label{sec:telep}

We restrict ourselves to two-level systems, and use the terminology of
spin-$\frac{1}{2}$ particles. The results can be applied to any representation
of qubits, e. g. for the polarization of single photon states, for which
quantum teleportation is experimentally feasible \cite{nature390_575}.  The
spin $z$-component eigenstates of the particles are denoted by
$|\uparrow\rangle$ and $|\downarrow\rangle$.  Density matrices of the
bipartite systems, if not otherwise stated, are represented on the natural
product state basis
$\{|\uparrow\rangle|\uparrow\rangle,|\uparrow\rangle|\downarrow\rangle,|\downarrow\rangle|\uparrow\rangle,|\downarrow\rangle|\downarrow\rangle
\}$. We will also use the notation
\begin{eqnarray}
  \label{eq:bellstates}
  |\Psi^{(\pm)}\rangle=\frac{1}{\sqrt{2}}\left(|\uparrow\rangle|\downarrow\rangle\pm|\downarrow\rangle|\uparrow\rangle \right)\nonumber \\
  |\Phi^{(\pm)}\rangle=\frac{1}{\sqrt{2}}\left(|\uparrow\rangle|\uparrow\rangle\pm|\downarrow\rangle|\downarrow\rangle \right)
\end{eqnarray}
for the four Bell-states.

Let us examine Bennett's scheme of quantum teleportation
\cite{prl70_1895} of a qubit: the sender, Alice has particle 1 of
$\frac{1}{2}$ spin in the state
\begin{equation}
  \label{eq:instate}
  \varrho^{(1)}_{\rm in}=\left( \matrix{ \varrho_{00}& \varrho_{01}\cr
                                     \varrho_{10}&
                                     \varrho_{11}}\right),
\end{equation}
and wants to teleport it to Bob. The upper indices of density matrices (and
other operators) refer to the number assigned to the particles.  They share
particles 2 and 3 in a state $\varrho^{(23)}$ as an entangled resource, and
there is also a classical communication channel between them. Alice has an
ideal Bell-state detector, and Bob can carry out unitary transformations on
particle 3 given to him.  As a starting point of our investigation, let us
suppose, that the state of particles 2 and 3 is described by the following
density matrix:
\begin{eqnarray}
  \label{eq:entangled}
  \varrho^{(23)}_{\rm shared}(\alpha)=\frac{1}{2}\bigl(
|\uparrow_2\rangle|\downarrow_3\rangle\langle\uparrow_2|\langle
\downarrow_3| +
|\downarrow_2\rangle|\uparrow_3\rangle\langle\downarrow_2|\langle
\uparrow_3| \nonumber \\
-\alpha |\uparrow_2\rangle|\downarrow_3\rangle\langle\downarrow_2|\langle
\uparrow_3| 
-\alpha|\downarrow_2\rangle|\uparrow_3\rangle\langle\uparrow_2|\langle \downarrow_3|
\bigr).
\end{eqnarray}
The lower indices in the kets indicate the number assigned to the particles.
Let $\alpha$ be real parameter between $0$ and $1$. For $\alpha=1$,
$\varrho^{(23)}_{\rm shared}(1)=|\Psi^{(-)}_{23}\rangle \langle
\Psi^{(-)}_{23}|$, thus in this case Alice and Bob share an EPR pair.  This is
the case of ideal quantum teleportation. Otherwise, the off-diagonal elements
of the density matrix are multiplied by $\alpha $.  For $\alpha=0$ the density
matrix is diagonal in the product state basis, describing classical statistics
only. This state is the mixture of the product states $|\uparrow_2\rangle
|\downarrow_3\rangle$ and $|\downarrow_2\rangle |\uparrow_3\rangle$.  Such a
state can be generated by a \emph{classically stochastic} source emitting
particles with opposite spins, with equal probability of sending ``up'' and
``down'' state both for particles 2 and 3.

The states in equation (\ref{eq:entangled}) may be rewritten in the
Bell-basis, we find
\begin{equation}
  \label{eq:entang_bellb}
  \varrho^{(23)}_{\rm shared}(\alpha)=
\frac{1+\alpha}{2}|\Psi^{(-)}_{23}\rangle\langle\Psi^{(-)}_{23}|
+\frac{1-\alpha}{2}|\Psi^{(+)}_{23}\rangle\langle\Psi^{(+)}_{23}|,
\end{equation}
that is, the state is a mixture of the two $\Psi$ Bell-states.  The class of
states examined here is a subclass of Werner states (which were examined from
teleportation's point of view in references~\cite{pla210_157,pra61_040101})
where the two $\Phi$ states do not occur. 

Let us consider now the entire teleportation process. Initially, the
state of the whole system of the three particles is the product of the
states in Eqs.~(\ref{eq:instate}) and (\ref{eq:entangled}):
\begin{equation}
  \label{eq:in123}
  \varrho_{\rm in}^{(123)}=\varrho_{\rm in}^{(1)}\otimes\varrho_{{\rm
  shared}}^{(23)}(\alpha).
\end{equation}
Alice carries out a Bell-state measurement, which is described by one
of the operators including projection of subsystems 1 and 2 to a Bell-state,
\begin{equation}
  \label{eq:bellmeas}
  \hat
  P_i^{(123)}= 4 \left(|\Psi_i^{(12)}\rangle\langle\Psi_i^{(12)}|
  \otimes \mbox{id}^{(3)}\right),
\end{equation}
where $|\Psi_{i}\rangle$ stands for one of the four Bell-states, and
$\mbox{id}^{(3)}$ is the identity operator for the Hilbert-space of particle
3. The result of the measurement is $i$, corresponding to the $i$-th
Bell-state. This information is sent to Bob via the classical channel. Because
the detection of each Bell-state occurs with probability $\frac{1}{4}$, the
operator is multiplied by $4$ in order to preserve the norm of the state
obtained.  The state of the system after the measurement is given by applying
the operator in equation~(\ref{eq:bellmeas}) to the state in
equation~(\ref{eq:in123}).  From this we obtain the state of Bob's particle by
tracing out in the other two particles:
\begin{equation}
  \label{eq:interm}
  \varrho_{\rm u}^{(3)}=\mbox{Tr}_{12}\left(P_i^{(123)}
\varrho_{\rm in}^{(123)}P_i^{(123)\dag}\right).
\end{equation}
In the last step, Bob has to apply a unitary transformation $U^{(3)}$ on state
$\varrho_{ u}^{(3)}$, according to Alice's measurement result $i$, which is
the identity operator in case Alice has detected $|\Psi^{(-)}\rangle$, and
\begin{equation}
  \label{eq:u}
  U^{(3)}=
\left(\matrix{ 1 & 0 \cr 0 & -1}\right),
\left(\matrix{ 0 & 1 \cr 1 & 0}\right),
\left(\matrix{ 0 & i \cr -i & 0}\right),
\end{equation}
for detecting $|\Psi^{(+)}\rangle$,$|\Phi^{(-)}\rangle$, and
$|\Phi^{(+)}\rangle$ respectively. Carrying out the calculations
described, we obtain our main result: the state teleported to Bob
reads
\begin{equation}
  \label{eq:result}
  \varrho^{(3)}_{\rm out}=\left( \matrix{ \varrho_{00}& \alpha \varrho_{01}\cr
                                     \alpha \varrho_{10}&
                                     \varrho_{11}}\right).
\end{equation}
It can be seen that \emph{the reduction of the off-diagonal elements
  of the density matrix describing the entangled resource is inherited by
  the teleported state.}
Thus teleportation acts as a phase-damping channel.

\section{The one-time-pad as a classical limit of teleportation}
\label{sec:class}

As a first example, let us examine the case of ``teleporting'' a classical
bit.  Assume, that $\alpha=0$, that is the density matrix describing the
entangled resource is diagonal. According to (\ref{eq:result}), only the
diagonal matrix elements of the density matrix of the input state, i.e. the
statistics of the measurement of the spin-$z$ component is
preserved. Therefore, let us suppose, that the input state in
equation~(\ref{eq:instate}) is already diagonal:
$\varrho_{10}=\varrho_{01}=0$. In this section we consider measurement of the
$z$ components of the spins.

Under these circumstances our particles can be exactly identified with
classical bits. The states of these classical bits, denoted by $\uparrow $,
and $\downarrow $, are identical with the basis quantum states
$|\uparrow\rangle$ and $|\downarrow\rangle$. The diagonal density matrix of
the input quantum state describes a classical probability distribution of bit
1.  This is transferred into bit 3 via a classical communication channel and a
\emph{classical} correlation. The process itself can be interpreted in the
following way: The source of bits 2 and 3 broadcasts correlated bit-pairs
$\downarrow_2\uparrow_3$ or $\uparrow_2\downarrow_3$ with equal probability.
This can be regarded as the classical limit of an EPR-pair. Bit 2 is obtained
by Alice, who makes a measurement, which tells whether bit 1 and 2 are the
same or different. One cannot speak of superpositions in this classical
context, and therefore the two $\Psi$ and the two $\Phi$ Bell-states coincide
in this limit: the former two mean simply ``the two bits are different''
($\Psi$-detection), and the latter ``the two bits are the same''
($\Phi$-detection). Thus the Bell-state measurement degenerates to an
``exclusive or'' operation, resulting in a single bit of information,
communicated to Bob.  Bob has to carry out the proper transformation to regain
the ``teleported'' bit.  There is no phase of the probability amplitudes for
classical probability, thus the transformations in equation~(\ref{eq:u})
degenerate to a conditional NOT operation: in case of ``$\Psi$ detection'',
Bob has obtained bit 3 in the proper state, while in case of ``$\Phi$
detection'' he has to invert bit 3.  Finally bit 3 is left with the original
value of bit 1.  Since the values of bits 1 or 2 are irrelevant from the
``Bell-state measurement'', it is not necessary for Alice to be aware of the
actual value of bit 1 to be ``teleported''. Therefore the method works for
``teleporting'' an unknown classical bit as well, similarly to the quantum
protocol.

The classical protocol described here is well known as the
one-time-pad in classical cryptography \cite{applcrypt}. This is, in
some sense, the classical protocol most similar to quantum
teleportation. It is the ``teleportation'' of the (optionally unknown)
state of a classical bit via a classical communication channel and a
classical correlation. The classical correlation provides noise in the
lack of the classical communication channel, while the classical
channel itself is useless for reconstruction of the result without the
correlation. Note that instead of measuring the state of bit 1 and
simply communicating its value, only a comparison with a reference has
been made.

\section{Statistics of a gedankenexperiment}
\label{sec:stat}

\begin{figure}
  \begin{center}
    \epsfbox{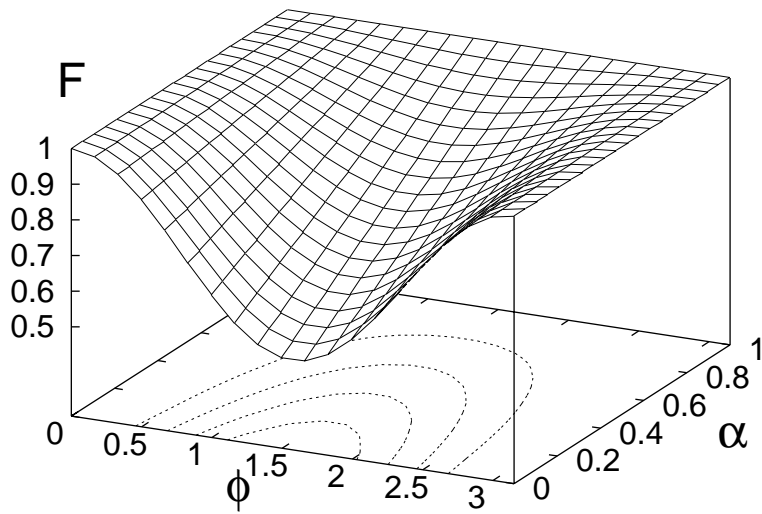}
  \end{center}
  \caption{The fidelity of the teleportation of the spin pointing to the direction described by the angle $\phi$ plotted against the parameter $\alpha$ describing the impurity of the state.}
  \label{fig:fig1}
\end{figure}

Having described the classical analogue, we may now examine $\alpha \neq 0$,
which interpolates between the classical and the quantum case. We calculate
the consequence of (\ref{eq:result}) to the result of a teleportation
experiment.  For simplicity, let us suppose, that we want to teleport the
state $|\uparrow \rangle$, having $z$-component of $+\frac{1}{2}$, rotated by
a given angle $\phi $ around the $x$ axis of the coordinate system.  The
operator of this rotation ($\hbar=1$) is
\begin{equation}
  \label{eq:rotop}
  \hat R(\phi)=e^{\frac{i}{2}\phi \sigma_x},
\end{equation}
$\sigma_x$ being the first Pauli-matrix,  and thus we
have the state
\begin{equation}
  \label{eq:inspin}
  |\Psi_{\rm in}(\phi)\rangle=\hat R(\phi)|\uparrow\rangle=
\left(\matrix{\cos(\frac{\phi}{2}) \cr i\sin(\frac{\phi}{2})}\right)
\end{equation}
to be teleported. According to (\ref{eq:result}), we obtain
\begin{equation}
  \label{eq:outspin}
  \varrho_{\rm out}(\phi,\alpha)=\left( \matrix{
\cos(\frac{\phi}{2})^2 &
-i \alpha\cos(\frac{\phi}{2})\sin(\frac{\phi}{2}) \cr
i \alpha\cos(\frac{\phi}{2})\sin(\frac{\phi}{2}) &
\sin(\frac{\phi}{2})^2 
}\right)
\end{equation}
as result of the teleportation process.  In order to verify the
teleportation, one may measure the spin along an axis obtained by
rotating the $z$ axis with the angle $\phi_m$ around the x axis.  The
probability of finding this component of the spin $+\frac{1}{2}$ is in
this case is
\begin{eqnarray}
  \label{eq:stat}
  \mathcal{P}(\phi,\phi_m,\alpha)=\mbox{Tr} \left(\varrho_{\rm out}
(\phi,\alpha)R(\phi_m)|\uparrow\rangle\langle\uparrow|R^\dag(\phi_m)\right)\nonumber\\
= \frac{\cos\left(\phi\right) \cos\left(\phi_m\right)}{2} + \frac{\alpha 
\sin\left(\phi\right) \sin\left(\phi_m\right)}{2} + 1/2.
\end{eqnarray}
For $\alpha=1$ we obtain the familiar cosine-type result valid for ideal
teleportation,
\begin{equation}
  \label{eq:stat_id}
  \mathcal{P}(\phi,\phi_m,1)=\frac{\cos(\phi-\phi_m)+1}{2},
\end{equation}
which is equal to $1$ for $\phi=\phi_m$, meaning perfect teleportation
to any direction.  For a given input state $|\Psi_{\rm in}(\phi)\rangle$,
the probability for finding the output in the input state after the
teleportation is the fidelity of the teleportation of the input state.
This fidelity is
\begin{equation}
  \label{eq:fidel}
  \mathcal{P}(\phi,\phi,\alpha)=-\frac{1}{4}\left[\alpha\left(\cos 2 \phi -
  1\right)-\cos 2 \phi - 3\right].
\end{equation}
In figure~\ref{fig:fig1} we have plotted this function. It is equal to
1 in the case of ideal teleportation. The basis states $|\uparrow
\rangle$ and $|\downarrow \rangle$ are always properly teleported, as
we have seen in the classical case. The minimum of the fidelity for
any $\alpha $ occurs for the equal superposition of the two basis
states ($\phi_m=\pi/2$). The minimum value for $\alpha=0$ is
$\frac{1}{2}$, expressing that the result of the measurement can be
either $\uparrow$ or $\downarrow$ with equal probability, thus this
state is not at all teleported.  Increasing the purity of the mixed
state increases the domain of the angles in which teleportation can be
regarded as reliable.

\section{Conclusion}
\label{sec:concl}

We have studied the role of the off-diagonal density matrix elements of the
entangled pair in the teleportation of a qubit.  By setting these to zero, we
have obtained the classical one-time pad cypher, as a classical limit of
quantum teleportation. We have demonstrated the result on the behavior of
statistics and fidelity of a gedankenexperiment. We have found, that the
fidelity of the teleportation is most sensitive to the loss of the quantum
entanglement in the EPR pair for a state at a right angle to the quantization
direction.

\section*{Acknowledgements}

This work was supported by the Research Fund of Hungary (OTKA) under contracts
Nos. T023777, T034484 and F032346. T. K. acknowledges the support of the
Hungarian Academy of Sciences (Bolyai J\'anos Kutat\'asi \"Oszt\"ond\' \i j).

\end{document}